\documentclass{aa}
\input epsf
\begin{document}

   \thesaurus{06     
       (02.01.2;		
	02.02.1;		
	08.02.3;		
	08.09.2 Cyg X-1;	
	13.25.3;		
	13.25.5;		
)}
   \title{Reflection and noise in Cygnus X--1}

   \author{M. Gilfanov\inst{1,2}, E. Churazov\inst{1,2},
	M. Revnivtsev\inst{2,1}  
	}

   \offprints{gilfanov@mpa-garching.mpg.de}

   \institute{Max-Planck-Institute f\"ur Astrophysik,
              Karl-Schwarzschild-Str. 1, D-85740 Garching bei M\"unchen,
              Germany
        \and
              Space Research Institute, Russian Academy of Sciences,
              Profsoyuznaya 84/32, 117810 Moscow, Russia
             }
  \date{Received 24 June 1999/ Accepted 31 August 1999}

	\titlerunning{Reflection and noise in Cygnus X--1}
	\authorrunning{M.Gilfanov et al.}
	
   \maketitle

   \begin{abstract}

We analyzed RXTE/PCA observations of Cyg X--1 from 1996--1998. 
We found a tight correlation between the characteristic noise
frequencies (e.g. the break frequency  $\nu_{\rm br}$) and the
spectral parameters in the low spectral state.
The amplitude of reflection increases and the spectrum of primary
radiation steepens as the noise frequency increases ($\nu_{\rm br}$
changes by a factor of $\sim 15$ in our sample).
This can be understood assuming that increase of the
noise frequency is associated with the shift of the inner boundary of
the optically thick accretion disk towards the compact object. 
The related increase of the solid angle, subtended by the
disk, and of the influx of the soft photons to the Comptonization
region lead to an increase of the amount of reflection and 
steepening of the Comptonized spectrum. The correlation between the
slope of primary radiation and the amplitude of reflection extends to
the soft spectral state. 

A similar correlation between reflection and slope was found for the
frequency resolved spectra in the 0.01-15 Hz frequency range.
Such a correlation could appear if the longer time scale variations
are associated with emission originating  closer to the optically 
thick disk and, therefore, having steeper Comptonized spectra with
larger reflection.

   \end{abstract}

%

\sloppypar

\section{Introduction}

Comptonization of soft seed photons in a hot, optically thin,
electron cloud located in the vicinity of a compact object is thought
to  form  the hard X--ray radiation in the
low spectral state of  Cyg X--1 (\cite{str79}). For a 
thermal distribution of the electrons with temperature $T_{\rm e}$, the
spectrum of the Comptonized radiation is close to a power law at 
energies sufficiently lower than $3 kT_{\rm e}$ (\cite{st80}).
The slope of the Comptonized spectrum is
governed by the ratio of the  
energy deposited into the electrons and the influx of the
soft radiation into the Comptonization region; the lower the
ratio the steeper the Comptonized spectrum
(e.g. \cite{st89}, \cite{feedback}, \cite{kemer}). 

The deviations from a single slope power law observed in the spectra
of X--ray binaries in the $\sim 5-30$ keV energy band are mainly due
to the reprocessing (e.g. reflection) of the primary Comptonized 
radiation by a cold medium located in the vicinity of the primary
source.  A  plausible candidate for the reflecting medium is the
optically thick accretion disk surrounding the inner region occupied 
by  hot optically thin plasma. The main observables of the emission,
reprocessed by a cold neutral medium, are the fluorescent K$_{\alpha}$
line of iron at 6.4 keV, iron K-edge at 7.1 keV  and a broad hump at $\sim   
20-30$ keV (\cite{bst}, ~ \cite{fab}). 
These signatures are indeed  observed in the spectra of
X--ray binaries. Their particular shape and relative amplitude,
however, depend on the geometry of the primary source and the
reprocessing medium and the abundance of heavy elements. And it is
affected by effects such as ionization and proper motion
(e.g. Keplerian motion in  the disk) of the reflector and general
relativity effects in the  vicinity of the compact object. 
The amplitude  of the reflection features is proportional to the solid
angle $\Omega_{\rm refl}$ subtended by the reflector.
An accurate estimate of  $\Omega_{\rm refl}$ is complicated
and the result is strongly dependent on the details of the spectral
model. However,  $\Omega_{\rm refl}$ is known to vary rather strongly from
source to source and, for a given source, from epoch to epoch
(e.g. \cite{done}, \cite{zdz2}).

\begin{table*}
\small
\caption{The list of the observations used for the analysis, the best-fit
parameters of the spectral approximation and the logarithmic frequency
shift. 
\label{fit}} 
\tabcolsep=0.19cm
\begin{tabular}{lccrcccccc}
\hline
ObsID & Date & Time, UT & Exp.$^1$ & $\alpha~^2$ &
$R\sim\Omega/2\pi~^3$ & $\sigma_{smo}~^4$ & EW $^5$ &
$\chi^2$/dof $^6$ & freq.shift $^7$\\
\hline
10235-01-01-00 & 12/02/96&12:45--13:54 &  3173& $1.821\pm 0.008$& $0.58\pm 0.04$& $0.79\pm 0.08$& $ 141 \pm 12$& 41.5/43& $ 0.66 \pm 0.034$  \\
10235-01-03-00 & 17/02/96&01:35--02:45 &  3193& $1.812\pm 0.008$& $0.54\pm 0.04$& $0.84\pm 0.07$& $ 141 \pm 12$& 35.0/43& $ 0.61 \pm 0.017$  \\
10236-01-01-02 & 17/12/96&06:04--11:43 & 10499& $1.754\pm 0.007$& $0.40\pm 0.02$& $0.50\pm 0.11$& $  97 \pm 11$& 17.7/37& $ 0.34 \pm 0.017$  \\
10236-01-01-020& 16/12/96&15:58--23:00 &  9982& $1.749\pm 0.007$& $0.38\pm 0.02$& $0.57\pm 0.11$& $ 101 \pm 12$& 22.3/37& $ 0.31 \pm 0.015$  \\
10236-01-01-03 & 17/12/96&12:45--13:25 &  2114& $1.754\pm 0.007$& $0.41\pm 0.02$& $0.48\pm 0.11$& $ 101 \pm 12$& 23.6/37& $ 0.33 \pm 0.009$   \\
10236-01-01-04 & 17/12/96&22:21--00:32 &  4928& $1.750\pm 0.007$& $0.38\pm 0.02$& $0.45\pm 0.13$& $  89 \pm 13$& 17.9/37& $ 0.35 \pm 0.012$  \\
10238-01-03-00 & 03/02/97&19:30--22:05 &  6441& $1.706\pm 0.007$& $0.29\pm 0.02$& $0.40\pm 0.12$& $  86 \pm 11$& 24.3/37& $ 0.11 \pm 0.004$ \\
10238-01-04-00 & 07/04/96&15:32--22:03 & 11394& $2.131\pm 0.029$& $1.11\pm 0.11$& $1.00\pm 0.04$& $ 283 \pm 20$& 47.7/40& $ 1.54 \pm 0.021$  \\
10238-01-05-00 & 31/03/96&02:30--06:36 &  2346& $1.900\pm 0.009$& $0.63\pm 0.04$& $0.86\pm 0.08$& $ 139 \pm 13$& 24.0/40& $ 1.00 \pm 0.023$  \\
10238-01-05-000& 30/03/96&19:54--01:58 & 12198& $1.913\pm 0.008$& $0.67\pm 0.04$& $0.86\pm 0.07$& $ 137 \pm 12$& 17.0/40& $ 1.00 \pm 0.007$ \\
10238-01-06-00 & 29/03/96&11:43--17:33 & 12064& $1.851\pm 0.007$& $0.52\pm 0.03$& $0.82\pm 0.07$& $ 146 \pm 12$& 34.0/40& $ 0.76 \pm 0.012$  \\
10238-01-07-00 & 28/03/96&05:48--11:05 &  9230& $1.833\pm 0.007$& $0.49\pm 0.03$& $0.85\pm 0.07$& $ 150 \pm 13$& 36.7/40& $ 0.70 \pm 0.011$  \\
10238-01-07-000& 27/03/96&23:06--05:20 & 10797& $1.860\pm 0.008$& $0.58\pm 0.04$& $0.92\pm 0.07$& $ 163 \pm 14$& 36.5/40& $ 0.76 \pm 0.019$  \\
10238-01-08-00 & 26/03/96&17:57--20:48 &  6239& $1.843\pm 0.007$& $0.54\pm 0.03$& $0.70\pm 0.07$& $ 130 \pm 10$& 24.2/40& $ 0.70 \pm 0.011$  \\
10238-01-08-000& 26/03/96&10:12--17:36 & 15233& $1.815\pm 0.007$& $0.48\pm 0.03$& $0.74\pm 0.07$& $ 118 \pm 11$& 20.6/40& $ 0.66 \pm 0.009$ \\
20175-01-01-00 & 25/06/97&11:59--14:37 &  4067& $1.933\pm 0.009$& $0.64\pm 0.04$& $0.79\pm 0.07$& $ 162 \pm 14$& 30.7/37& $ 1.09 \pm 0.015$  \\
20175-01-02-00 & 04/10/97&11:45--13:49 &  4301& $1.753\pm 0.007$& $0.37\pm 0.02$& $0.41\pm 0.12$& $  87 \pm 12$& 19.3/37& $ 0.34 \pm 0.020$  \\
30157-01-01-00 & 11/12/97&05:23--06:37 &  3316& $1.750\pm 0.007$& $0.36\pm 0.02$& $0.35\pm 0.09$& $  80 \pm 10$& 24.9/37& $ 0.29 \pm 0.008$ \\
30157-01-03-00 & 24/12/97&21:29--23:03 &  2861& $1.707\pm 0.007$& $0.33\pm 0.02$& $0.34\pm 0.12$& $  74 \pm 10$& 14.1/37& $ 0.18 \pm 0.005$ \\
30157-01-04-00 & 30/12/97&18:18--19:00 &  2323& $1.723\pm 0.007$& $0.32\pm 0.02$& $0.36\pm 0.09$& $  86 \pm 10$& 25.8/37& $ 0.21 \pm 0.011$   \\
30157-01-05-00 & 08/01/98&04:45--06:29 &  3648& $1.744\pm 0.007$& $0.37\pm 0.02$& $0.33\pm 0.09$& $  83 \pm 10$& 18.5/37& $ 0.24 \pm 0.029$   \\
30157-01-07-00 & 23/01/98&01:06--02:01 &  2772& $1.753\pm 0.007$& $0.39\pm 0.02$& $0.37\pm 0.11$& $  84 \pm 11$& 26.3/37& $ 0.24 \pm 0.010$  \\
30157-01-08-00 & 30/01/98&01:10--02:24 &  2895& $1.755\pm 0.007$& $0.39\pm 0.02$& $0.28\pm 0.08$& $  74 \pm  9$& 20.5/37& $ 0.34 \pm 0.016$  \\
30157-01-09-00 & 06/02/98&00:02--00:34 &  1331& $1.747\pm 0.007$& $0.42\pm 0.02$& $0.46\pm 0.10$& $  71 \pm  9$& 17.5/37& $ 0.26 \pm 0.026$  \\
30157-01-09-01 & 05/02/98&23:32--00:01 &  1647& $1.758\pm 0.007$& $0.43\pm 0.02$& $0.39\pm 0.10$& $  74 \pm 10$& 15.5/37& $ 0.29 \pm 0.013$  \\
30157-01-10-00 & 13/02/98&23:31--00:28 &  2613& $1.751\pm 0.007$& $0.35\pm 0.02$& $0.34\pm 0.10$& $  76 \pm 10$& 20.6/37& $ 0.32 \pm 0.011$  \\
\hline
\end{tabular}\\
The energy spectra were fitted in the 4--20 keV energy range (see the
text for the details of the spectral model). 
The errors are $1\sigma$ for one parameter of interest.
$^1$ -- dead time corrected exposure time, sec;
$^2$ -- the power law photon index;
$^3$ -- the reflection scaling factor;
$^4$ -- the width of the Gaussian used to model smearing of the reflection
features, keV; 
$^5$ -- the equivalent width of the 6.4 keV line, eV;
$^6$ -- the $\chi^2$/dof of the spectral fit;
$^7$ -- the logarithmic frequency shift of the template power
spectrum, characterizing the  noise frequency.
\end{table*}

Based on the analysis of a large sample of Seyfert AGNs and several
X--ray binaries \cite{zdz1} recently found a correlation between the
amount of reflection and the slope of the underlying power law.  
They concluded that the existence of such a correlation implies a dominant
role of the reflecting medium as the source of seed photons to the
Comptonization region.

The power density spectra of X--ray binaries in the low state (see
\cite{qporev} for a review) are dominated by a band
limited noise component which is relatively flat below a break
frequency $\nu_{\rm br}$ and approximately 
follows a power law with a slope of $\sim 1$ or steeper above
the $\nu_{\rm br}$. 
Superimposed on the band limited noise is a broad bump,
sometimes having the appearance of a QPO, which frequency, $\nu_{\rm QPO}$,
is by an $\sim$order of magnitude higher than the break frequency of
the band limited noise. Although a number of theoretical models was
proposed to explain the power spectra of X--ray binaries
(e.g. \cite{alpar}, \cite{nowak}, \cite{ipser}), the
nature of the characteristic noise frequencies is still unclear.  
Despite of the fact that the characteristic noise frequencies 
$\nu_{\rm br}$ and $\nu_{\rm QPO}$ vary from source to source and from epoch
to epoch,  they follow a rather tight correlation,
$\nu_{\rm QPO}\propto\nu_{\rm br}^{\alpha}$, in a wide range of source
types and luminosities (\cite{br_qpo}, \cite{br_qpo1}). 

Analyzing the GRANAT/SIGMA data Kuznetsov et al. (1996, 1997) found a
correlation between 
the rms of aperiodic variability in a broad frequency range and the
hardness of the energy spectrum above 35 keV for Cyg X--1 and GRO
J0422+32 (X--ray Nova Persei). \cite{cyg_batse} came to similar
conclusions based on the data of CGRO/BATSE observations of Cyg X--1. 

We present below the results of a systematic analysis of the RXTE
observations of Cyg X--1 performed from 1996-1998 aimed to search for a
relation between characteristic noise frequencies and spectral
properties.

\begin{figure*}[t]
\hbox{	
\epsfxsize 9 cm
\epsffile{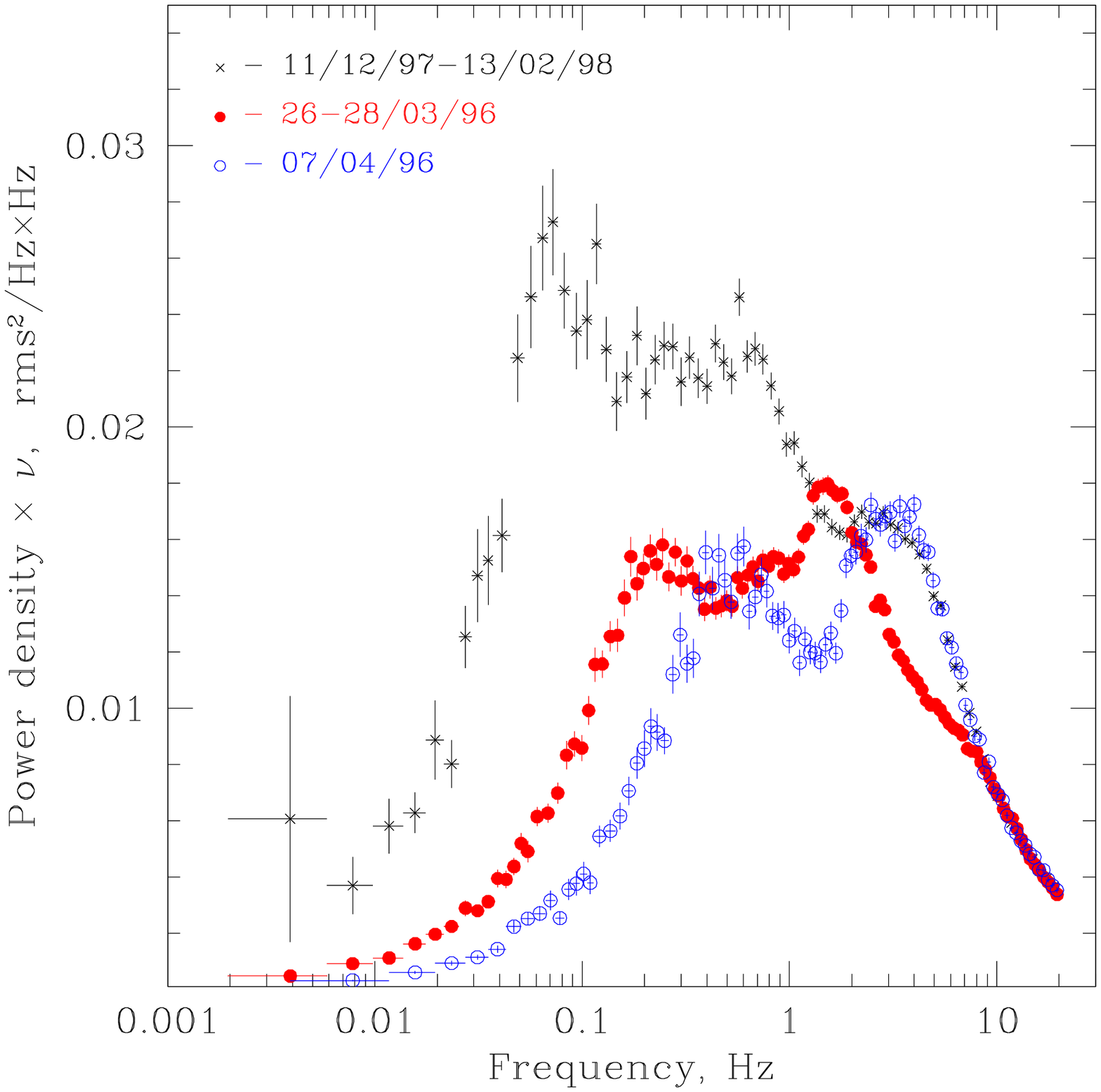}
\epsfxsize 9 cm
\epsffile{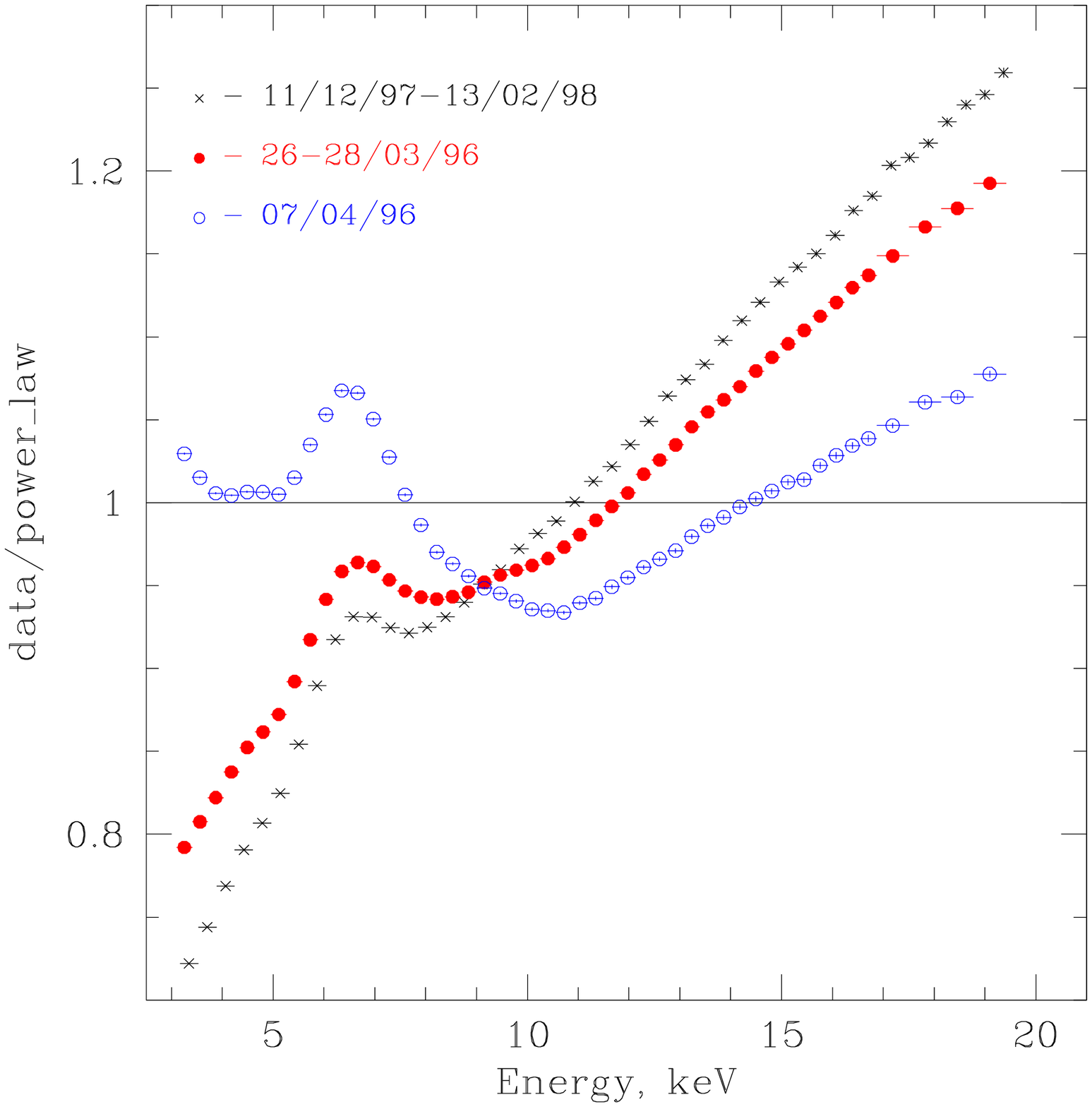}
}
\caption{ The power density ({\em left}) and counts ({\em right})
spectra of Cyg X--1 at different epoch. 
The power density spectra are plotted as frequency $\times$ (power
density), i.e. in units of $\rm Hz\times rms^2/Hz$.
The counts spectra are shown as a ratio to a single slope power law
model.
The symbols are the same in the left and right panels.
The higher characteristic noise frequencies correspond to steeper
energy spectra with stronger reflection features.
\label{pds_spe}
}
\end{figure*}

\section{Observations and data analysis}

We used the publicly available data of
Cyg X--1 observations with the Proportional Counter Array aboard the
Rossi X-ray Timing Explorer  performed between  
Feb. 1996 and Feb. 1998 during the low (hard) spectral state of the
source. In total our sample contained 26 observations randomly chosen
from proposals 10235, 10236, 10238, 20175 and 30157 (Table \ref{fit}). 
The energy and power density spectra were averaged for each individual
observation.  The 4-20 keV flux from Cyg X--1 varied from $\sim
7.2\cdot 10^{-9}$ to $\sim 1.8\cdot 10^{-8}$ erg/sec/cm$^2$ which
corresponds to luminosity   range of $\sim
5.4\cdot10^{36} - 1.3\cdot 10^{37}$ erg/sec assuming a 2.5 kpc
distance.

The power  spectra were calculated in the 2--16 keV energy
band and the $\approx 0.002-32$ Hz frequency range  following the
standard X--ray timing technique and normalized to units of squared
relative rms. The white noise level was corrected for the dead time
effects following \cite{qpomod} and \cite{zhang}.

The energy spectra were extracted from the ``Standard Mode 2'' data and
ARF and RMF were constructed using standard RXTE FTOOLS v.4.2 tasks.
We assumed a 0.5\% systematic error in the spectral fitting. 
The ``Q6'' model was used for the background calculation.
We used XSPEC v.10.0 (\cite{xspec}) for the
spectral fitting.

\section{Results}

\begin{figure}
\epsfxsize 9 cm
\epsffile{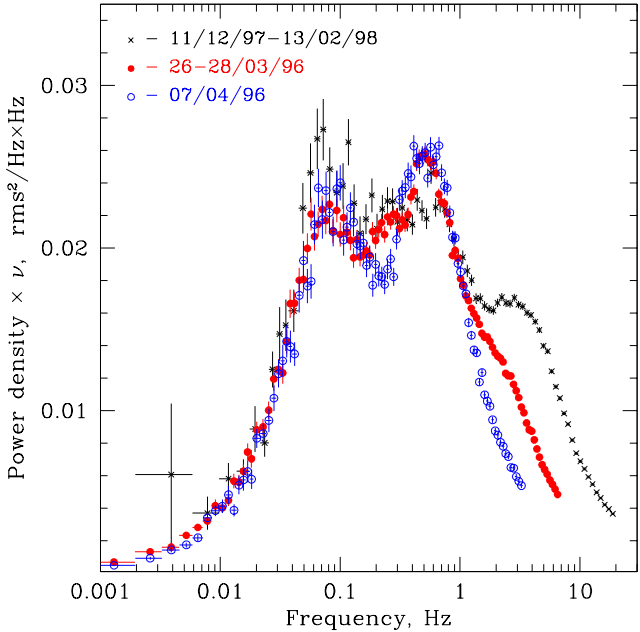}
\caption{The power density spectra of Cyg X--1 for the same datasets
as in Fig.\ref{pds_spe} but logarithmically
shifted along the frequency axis and renormalized to match the
spectrum averaged over 11/12/97--13/02/98 (thin black crosses) at low
frequencies. 
The power density spectra are plotted as frequency $\times$ (power
density), i.e. in units of $\rm Hz\times rms^2/Hz$.
The symbols are the same as in Fig.\ref{pds_spe}.
\label{pds_rescale}
}
\end{figure}

Several power density and counts spectra of Cyg X--1 observed at
different epoch are shown in Fig.\ref{pds_spe}. The power
density is plotted in the units of frequency$\times$power presenting
squared fractional rms at a given frequency per decade of
frequency. This way of representing the power spectra  clearly
characterizes  the relative contribution of variations at different
frequencies to the total observed rms. 
Most of the power of aperiodic variations  below $\sim 30$ Hz is
approximately  equally divided between two  broad, $\Delta\nu/\nu\sim
1$ ``humps'' separated in frequency  by a $\sim$decade
(the left panel in Fig.\ref{pds_spe}).  The lower frequency
``hump'' corresponds to what is  usually referred to as a band
limited noise (e.g. \cite{qporev}). The second ``hump'' is sometimes
called a ``QPO''. Both frequencies are equally important
quantitative characteristics of the aperiodic variability -- most of
the apparently observed variability, indeed, occurs roughly at these
two characteristic frequencies (Fig.\ref{pds_spe}).   
Despite of a factor of $\sim 10-15$ change of the characteristic noise
frequencies in our sample, the high frequency part of the power
spectrum, above $\sim 8$ Hz, remains unchanged (cf. \cite{belhas}).
However, the low and intermediate frequency parts, 
responsible for the most of the observed
variability, change in an approximately self--similar manner which
can be described as a logarithmic shift along the frequency axis
(Fig.\ref{pds_rescale}) (cf. \cite{pds_batse}).   
This is a manifestation of the fundamental correlation between the
break and the QPO frequencies found by Wijnands \& van der Klis (1999)
and Psaltis et al. (1999) for a broad range of the source types.

\begin{figure}
\hbox{	
\epsfxsize 9 cm
\epsffile{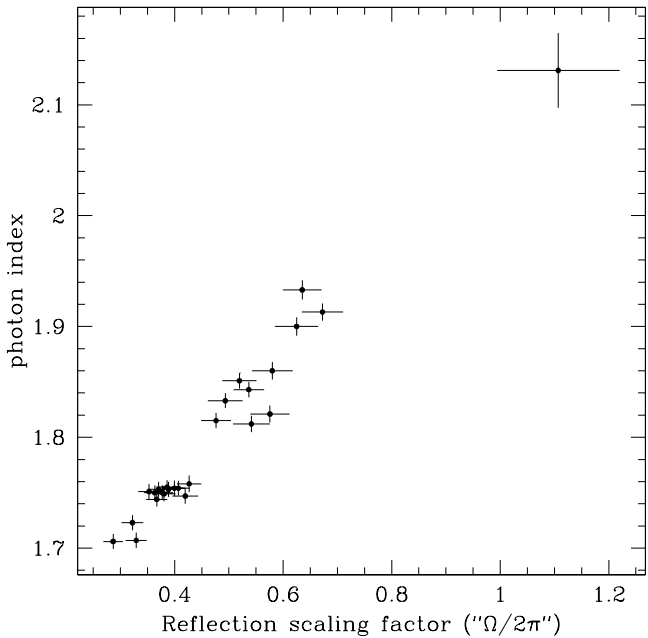}
}
\caption{The photon index of the underlying power law plotted
vs. reflection scaling factor. See text for discussion of the spectral
model. 
\label{slope}
}
\end{figure}

The counts spectra change in accordance with the change of the noise
frequency (cf. left and right panels in Fig.\ref{pds_spe}).
The increase of characteristic  noise frequency is accompanied by the
general steepening of the energy spectrum and an increase of the relative
amplitude of the reflection features.  

In order to quantify this effect we fit the energy spectra in our
sample with a simple model consisting of a power law with a
superimposed reflected continuum (pexrav model in XSPEC) and an
intrinsically narrow line at 6.4 keV. The binary system inclination
was fixed at $i=50\degr$ (\cite{incl2}, \cite{done}; see, however,
\cite{incl1}), 
the iron abundance was fixed at the solar value of $A_{\rm Fe}=3.3\cdot
10^{-5}$ and the low energy 
absorption at $N_{\rm H}=6\cdot 10^{21}$ cm$^{-2}$. Effects of ionization
were not included. 
In order to approximately include in the model the smearing of the 
reflection features due to motion in the accretion flow 
the reflection continuum and line were convolved with a gaussian,
which width was a free parameter of the fit. 
The spectra were fitted in the 4--20 keV energy range. 
For the spectrum of 07/04/96 (observation 10238--01--04--00)
an additional soft component (XSPEC diskbb model) was included in the
model. This observation occurred shortly before the soft 1996  state of
the source.  The energy spectrum had the largest amplitude of
reflection and the power spectrum showed the highest noise frequencies  
(open circles in Figs.\ref{pds_spe} and \ref{pds_rescale}). The
best--fit parameters are listed in Table \ref{fit}. 
The accuracy of the absolute values of the best--fit parameters is
discussed in the next section.

The model describes all the spectra in our sample with an accuracy
better than $\approx 1\%$ and reduced $\chi^2_{\rm r}$ in the range 
0.5-1.1.  
The quality of the fit decreases by somewhat with the increase of the 
best fit value of the reflection scaling factor. The 
model, however, fails to reproduce the exact shape of the line and, 
especially, of the edge.  Nearly the same pattern of systematic
deviations of the data from the model was found for all spectra in the
$\sim 6-12$ keV energy range. Inclusion in the model of the effect of
ionization and/or account for the exact shape of the relativistic
smearing, assuming Keplerian motion in the disk, does not change significantly
the pattern of residuals.

We found a fairly good correlation between the photon index
of the  underlying power law and the reflection scaling factor
$R\approx\Omega/2\pi$ roughly characterizing the solid angle subtended
by the  reflecting media (Fig.\ref{slope}). Other parameters
change in an anticipated way indicating that the the spectral  model
includes the most physically important features. 
In particular, the width of the Gaussian used to model the smearing of
the reflection spectrum and the equivalent width of the 6.4 keV line
increase as the reflection scaling factor increases, the 
equivalent width being roughly proportional to $R$.

\begin{figure}
\vbox{	
\epsfxsize 8.8 cm
\epsffile{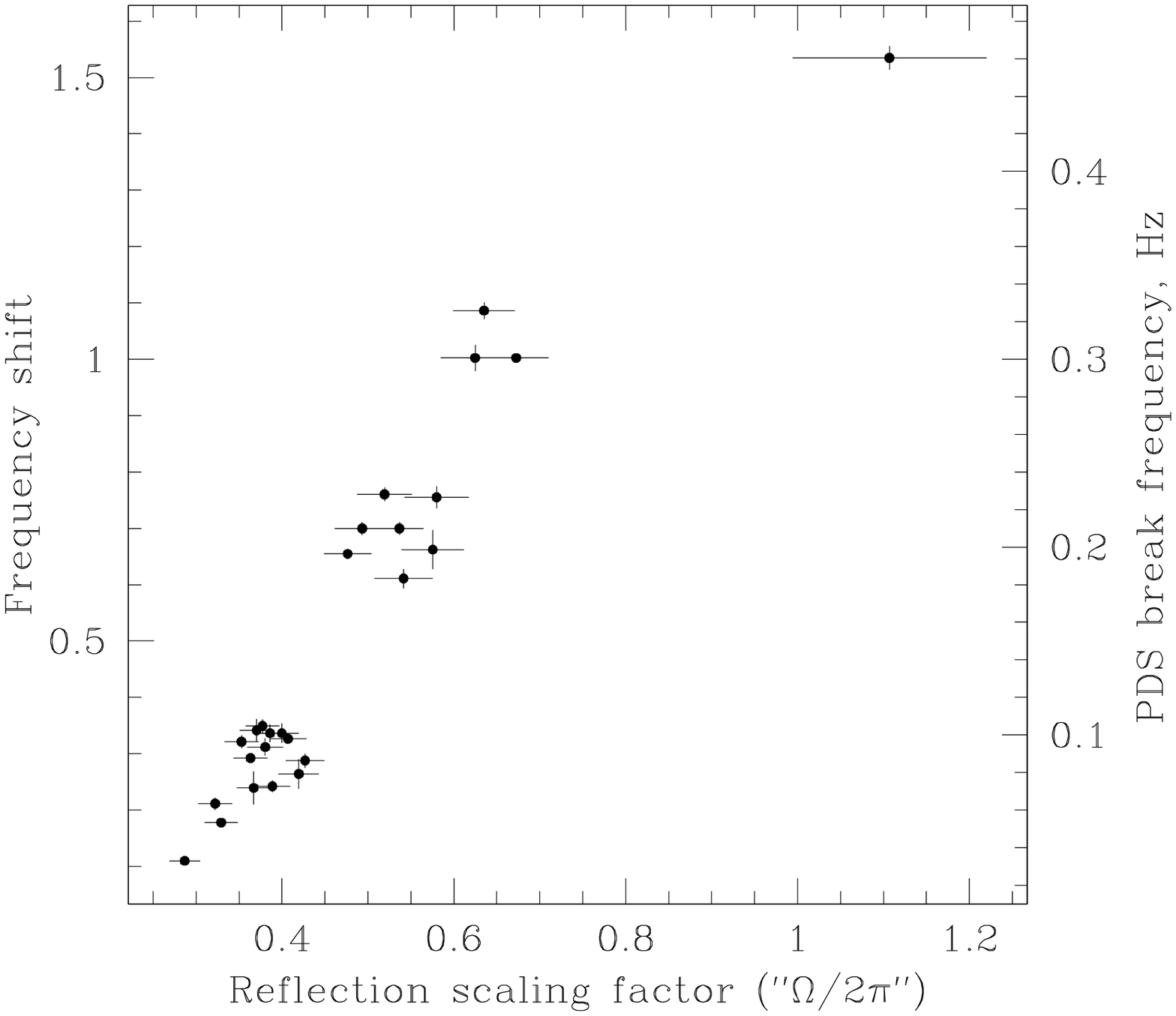}
\vspace{0.3cm}
\epsfxsize 8.8 cm
\epsffile{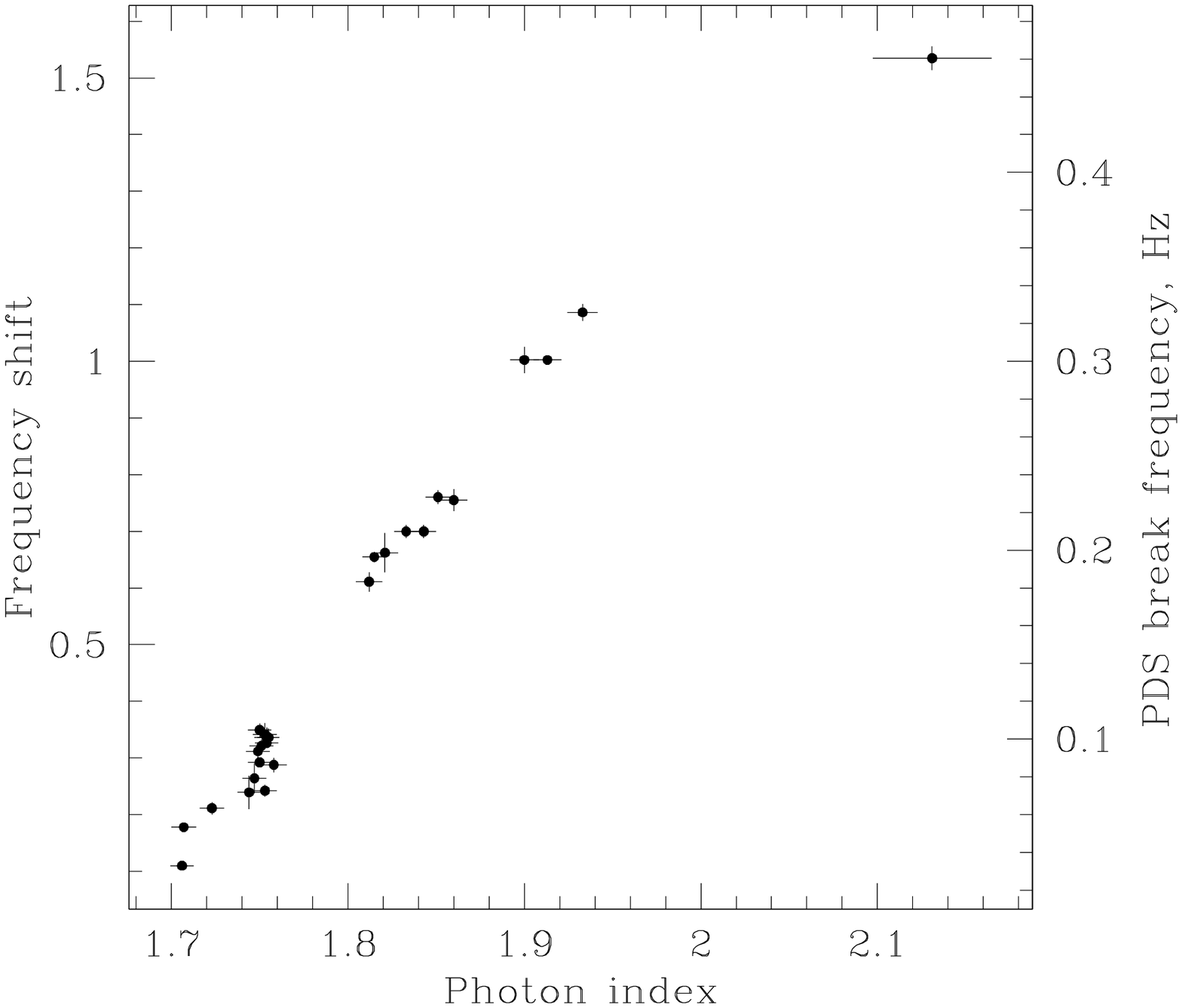}
}
\caption{The PDS frequency scaling factor plotted vs. reflection
scaling factor ({\em top}) and photon index of the underlying power
law ({\em bottom}). The vertical axis on the right hand side of the
plots is labeled in units of the PDS break frequency.
\label{shift}
}
\end{figure}

In order to relate spectral properties with characteristic noise
frequency we fit the power density spectra with a template, 
allowing logarithmic shift of the template spectrum  along the
frequency axis and its renormalization. Only the low and intermediate
frequency part of the power spectra, up to the second ``hump'', was
used in the fitting (cf. Fig.\ref{pds_rescale}). The best fit values
of the two parameter of the fit, the 
frequency shift and the normalization, were searched for using the
$\chi^2$ minimization technique. In order to compute the 
$\chi^2$, the template spectrum, logarithmically shifted along the
frequency axis, was rebinned to the original frequency bins using
linear interpolation . We chose the average power spectrum for 
observations 10238--01--05--00 and 10238--01--05--000 as a template.  
The best--fit values of the frequency scaling factor  obtained in such
a way are listed in Table \ref{fit}. Fig.\ref{shift} shows the
frequency scaling factor  plotted versus reflection amplitude (top)
and photon index of the underlying power law (bottom).

The change of the template power spectrum does not affect the correlations
shown in Fig.\ref{shift}. More sophisticated spectral models including
the effect of ionization and/or the exact shape of the relativistic
smearing, assuming Keplerian motion in the disk, change the particular
values of the best fit parameter but do not remove the general trends
in Figs. \ref{slope} and \ref{shift}. 
The same is true for a reflection model based on the results of our 
Monte--Carlo calculations of the reflected spectrum (an isotropic
source above an optically thick slab with solar abundance of heavy
elements; ionization effects included) in which the
equivalent width of the K--$\alpha$ line is linked to the amplitude of
the absorption K--edge and of the reflected continuum.
More sophisticated spectral models affect mostly the value of the
reflection scaling factor, increasing  the scatter of the points along
the horizontal axis in Fig.\ref{slope} and in the top panel in
Fig.\ref{shift}. The scatter of the values of photon index, on the
other hand, almost does not change.

\section{Discussion}

We found a correlation between the noise frequency and spectral
parameters, in particular, the  amount of reflection and the slope of
the underlying power law. The increase of the noise frequency is 
accompanied by the steepening of the spectrum of the primary radiation
and the increase of the amount of reflection.

The correlation between the spectral parameters -- amount of
reflection and the slope of the primary emission -- is the same as
recently found by \cite{zdz1} for a large sample of Seyfert AGNs and
several X--ray binaries.  
The existence of such a correlation hints at a close relation
between the solid angle subtended by the reflecting media and
the influx of the soft photons to the Comptonization region. More
specifically, it suggests that the reflecting media gives a dominant
contribution to the influx of the soft photons to the Comptonization
region. 
The geometry, commonly discussed in application to the low spectral state
of X--ray binaries, involves a hot quasi-spherical  Comptonization
region near the compact object surrounded by an optically thick
accretion disk. In such a geometry it is natural to expect that the
decrease of the inner radius of the disk would result
in an increase of the solid angle, subtended by the reflector (disk),
and an increase of the energy flux of the soft photons to the
Comptonization  region. The correlated behavior of the noise frequency
and spectral parameters suggests, that a decrease of the inner radius
of the disk leads also to an increase of the noise frequency.

\begin{figure}
\epsfxsize 9 cm
\epsffile{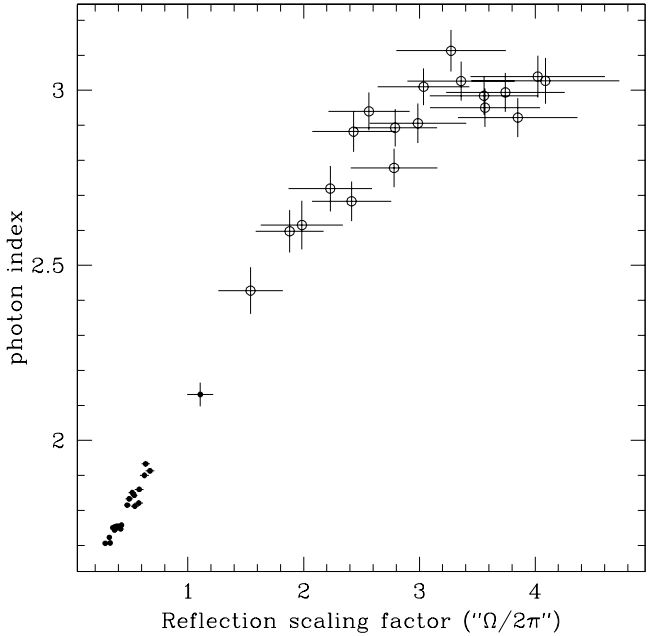}
\caption{The photon index of the underlying power law plotted
vs. reflection scaling factor for both low (solid circles) and high
(open circles) spectral states. The low state data are the same as in
Fig.\ref{slope}. The high state spectra were fit using the same
spectral model as the low state data with addition of a soft
multicolor disk component. The particular values of the reflection
scaling factor, especially for the soft state, are subject to a number
of uncertainties (see discussion in the text and Figs.\ref{low_high},
\ref{sprat}).
\label{slope_high}
}
\end{figure}

\begin{figure}[t]
\epsfxsize 9 cm
\epsffile{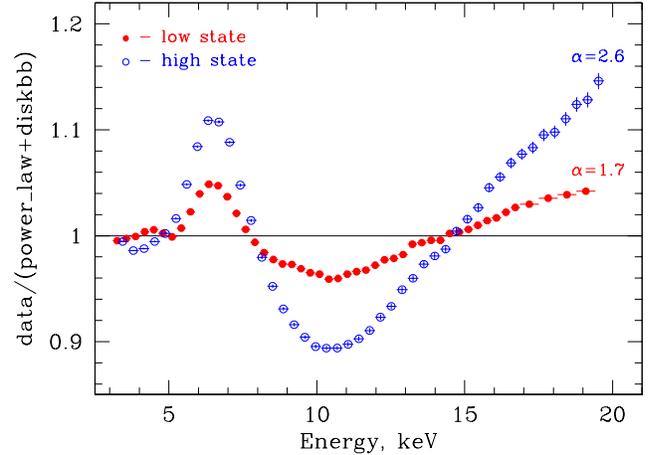}
\caption{The ratio of the counts spectra to a power law + multicolor
blackbody disk emission model for a typical low state spectrum and a
soft state spectrum. Each ratio is
labeled by the power law photon index used to calculate the ratio.
\label{low_high}
}
\end{figure}

In the soft state the inner boundary of the optically
thick disk is likely to shift  closer to the compact object,
$R_{\rm d}\sim 3R_{\rm g} $ (cf. $R_{\rm d}\sim 15-100 R_{\rm g}$ in
the hard state, \cite{done}, \cite{texas96}). Correspondingly,  one
might expect that the  
soft state spectra should have stronger reflected component.
An accurate estimate of the spectral parameters in the soft state
is a complicated task and is beyond the scope of this paper.
However, in order to qualitatively check this hypothesis we analyzed
a set of RXTE observations of Cyg X--1 in the soft spectral state
(May--August 1996). The spectral model was identical to the one used
for the  analysis of the low state data with addition of a
disk component (diskbb model in XSPEC); the energy range was 3--20
keV. We found
that  the correlation between the slope of the primary emission and
the amount of reflection continues smoothly into the soft state 
(Fig.\ref{slope_high}), but the best--fit values of the reflection
scaling factor are too high and, in particular, considerably
exceed unity.
However the qualitative conclusion that the amount of reflection increases
from the low to the high spectral state is evident from the comparison of
typical low and high state spectra (Fig.\ref{low_high}). 
The results of Done \& Zycki (1999) and Gierlinski et al. (1999)
based on more
realistic spectral models also  confirm the existence of such a trend --
$\Omega/2\pi\sim 0.1$ and $\Omega/2\pi\sim 0.6-0.7$ for the low and the
high state respectively.

\begin{figure}
\epsfxsize 9 cm
\epsffile{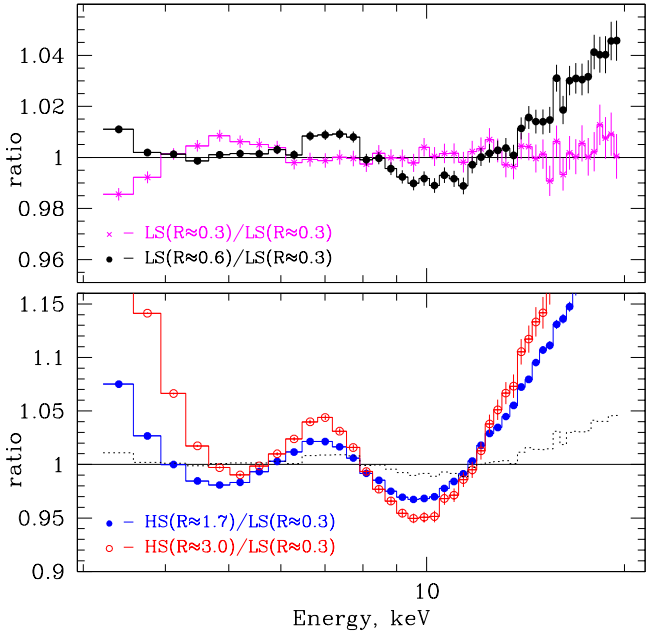}
\caption{The ratio of the counts spectra in the low (the upper panel)
and high (the lower panel) spectral states with different best--fit
value of the reflection scaling factor to the low state spectrum
with lowest reflection ($R\approx 0.3$). The ratios are multiplied by
different power law functions of energy and renormalized.
The high and low spectral states are
denoted in the legend as ``LS'' and ``HS'' respectively. The best--fit
values of the reflection scaling factor for different spectra are
indicated in the legend. The dotted line in the lower panel shows the 
ratio for a low state spectrum with  $R\approx 0.6$ (the same spectrum
as in the upper panel). 
\label{sprat}
}
\end{figure}

The spectral model is obviously oversimplified. 
Therefore the best fit values  do not necessarily  
represent the exact values of the physically meaningful parameters. 
Particularly subject to the uncertainties due to the choice of the
spectral model are the reflection scaling factor $R\sim\Omega/2\pi$
and the  equivalent width of the iron line.
Our estimates of the  reflection scaling factor 
for the low state are systematically higher than those
typically obtained   using the more elaborate models $\Omega/2\pi\sim
0.1-0.2$  (e.g. \cite{done}). Moreover, the best--fit values of the 
$R\sim\Omega/2\pi$ for the high spectral state exceed the unity
considerably, what is implausible in the usually adopted geometry of
the accretion flow.
More realistic models, however, impose stringent
requirements on the quality and energy range of the data in order
to eliminate the degeneracy of the parameters. We therefore chose a model
including the most  physically  important features and satisfactorily
describing the data, and on the other hand, having a minimal number of
free, especially, mutually dependent parameters. Although the absolute
values of the best--fit parameters obtained with such a model should
be treated with  caution, the model correctly ranks the spectra
according to the importance of the reprocessed component. In order to
demonstrate this we plotted in Fig.\ref{sprat}  the ratio of several
counts spectra in the low and the high state with different best--fit
values of $R\sim\Omega/2\pi$ to the spectrum with the lowest value of
reflection in our sample. The Fig.\ref{sprat} clearly shows that the
spectra having higher best-fit values of $R\sim\Omega/2\pi$ have
more pronounced reflection signatures -- the fluorescent K$_{\alpha}$ 
line of iron at $\sim 6-7$ keV and broad smeared iron K-edge at 
$\sim 7-10$ keV. Similarly we used  
a simple way of quantifying the characteristic noise frequency in
terms of a logarithmic shift of a template spectrum along the
frequency axis.

\begin{figure}
\vbox{	
\epsfxsize 9 cm
\epsffile{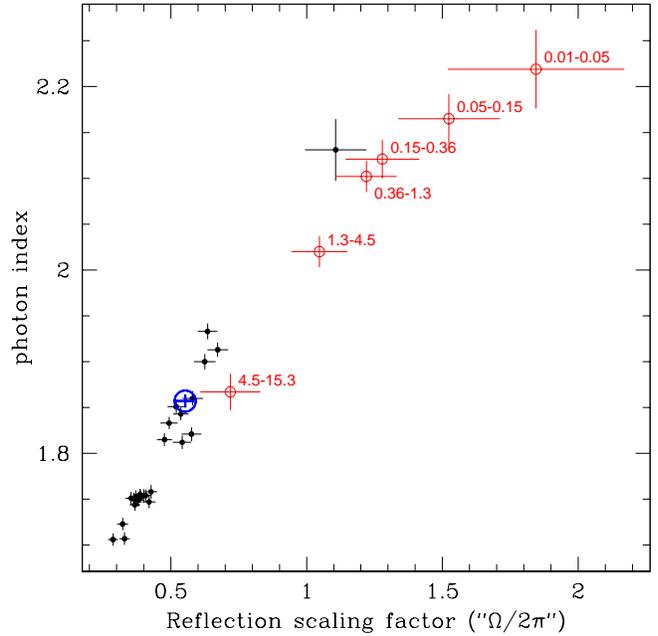}

}
\caption{The photon index of the underlying power law plotted
vs. the reflection scaling factor for the frequency resolved and average
spectra. Solid circles show the best fit parameters 
for the averaged spectra from our sample (the same data as in
Fig.\ref{slope}). The open circles are best fits to the frequency
resolved spectra. The numbers near the error bars indicate the
frequency range in Hz. The large open circle is a best fit to the
spectrum averaged over the datasets used to calculate the frequency
resolved spectra. 
\label{slope_frsp}
}
\end{figure}

Recently, \cite{mikej1} applied  a frequency resolved spectral
analysis  to the data of Cyg X--1 observations. They
showed that energy spectra corresponding to the shorter time scales
($\la 0.1-1$ sec) exhibit less reflection than that of the longer time
scales. 
Interpretation of the frequency resolved spectra is not
straightforward and requires some a priori assumptions. 
We shall assume below that the different time scale variations occur
in geometrically distinct regions of the accretion flow and the
spectral shape does not change during a variability cycle on a given
time scale. Under these assumptions the frequency resolved spectra can
be treated as representing the energy spectra of the events occurring
on the different time scales.  
We reanalyzed the  data from Revnivtsev et al. (1999)
using the spectral model described in the previous section. 
We found that the frequency resolved spectra 
follow  the same trend as the averaged energy spectra
(Fig.\ref{slope_frsp}), thus confirming the existence of an intimate
relation between the slope of primary radiation and the amount of
reflection.  Secondly, energy spectra of the longer time scale ($\sim
0.01-5$ Hz) variations, giving the dominant contribution to the 
observed rms, are considerably softer and contain more reflection than
the averaged energy spectrum. 
Such behavior  hints at the non-uniformity of the conditions
in the Comptonization region. 
Higher frequency variations are associated with a (presumably
inner) part of the Comptonization region having a smaller solid angle,
subtended by the disk, and a larger ratio of the energy deposited into 
the electrons to the flux of soft seed photons from the disk.

\section{Conclusions}

We analyzed a number of RXTE/PCA observations of Cyg X-1 from
1996--1998. 
\begin{enumerate}
\item We found a tight correlation between characteristic noise
frequency and spectral parameters -- the slope of primary Comptonized
radiation and the amount of reflection in the low spectral state
(Fig.\ref{slope}, \ref{shift}).  
We argue that the simultaneous increase of the noise frequency, the
amount of reflection and the steepening of the spectrum of the
Comptonized radiation are caused by a decrease of the inner radius of
the optically thick accretion disk. 
\item The soft state spectra have larger reflection than the low state
spectra  and obey the same correlation between the slope of the
Comptonized radiation and the amount of reflection
(Fig.\ref{slope_high}). 
\item 
A similar correlation between the slope of the primary radiation and
the amount of reflection was found for the frequency resolved
spectra. 
The energy spectra at the lower frequencies (below $\sim$
several Hz), responsible for most of the apparently
observed aperiodic variability,  are considerably steeper and contain
a larger amount of reflection than the spectra of the higher frequencies 
and, most importantly, than the average spectrum. We
suggest that this reflects non-uniformity and/or non-stationarity of
the conditions in the Comptonization region. 

\end{enumerate}

\begin{acknowledgements}
The authors are grateful to R.Sunyaev for stimulating discussions
and valuable comments on the manuscript. M.Revnivtsev acknowledges
hospitality of the Max--Planck Institute for Astrophysics and a partial
support by RBRF grant 96--15--96343 and INTAS grant 93--3364--ext.
This research has made use of data obtained through the High Energy
Astrophysics Science Archive Research Center Online Service, provided 
by the NASA/Goddard Space Flight Center.
\end{acknowledgements}

\end{document}